\documentclass{kapproc} 

\let\footnotetext\savefootnotetext

\let\footnoterule\savefootnoterule

\normallatexbib

\begin{document}

\def\d{\partial}   
\def\str{{\rm STr \;}}
\def\({\left(}     
\def\){\right)}
\def\dx{ \frac{d^2 x}{2\pi} }
\def\vep{\varepsilon}
\def\psib{\bar \psi}
\def\zbar{{\bar z}}
\def\inv#1{{1 \over #1}}
\def\betab{\bar{\beta}}
\def\half{\frac{1}{2}}
%
\def\th{\theta}        \def\Th{\Theta}
\def\ga{\gamma}        \def\Ga{\Gamma}
\def\be{\beta}
\def\al{\alpha}
\def\ep{\epsilon}
\def\vep{\varepsilon}
\def\la{\lambda}    \def\La{\Lambda}
\def\de{\delta}        \def\De{\Delta}
\def\om{\omega}        \def\Om{\Omega}
\def\sig{\sigma}    \def\Sig{\Sigma}
\def\vphi{\varphi}
%
%
\def\CA{{\cal A}}    \def\CB{{\cal B}}    \def\CC{{\cal C}}
\def\CD{{\cal D}}    \def\CE{{\cal E}}    \def\CF{{\cal F}}
\def\CG{{\cal G}}    \def\CH{{\cal H}}    \def\CI{{\cal J}}
\def\CJ{{\cal J}}    \def\CK{{\cal K}}    \def\CL{{\cal L}}
\def\CM{{\cal M}}    \def\CN{{\cal N}}    \def\CO{{\cal O}}
\def\CP{{\cal P}}    \def\CQ{{\cal Q}}    \def\CR{{\cal R}}
\def\CS{{\cal S}}    \def\CT{{\cal T}}    \def\CU{{\cal U}}
\def\CV{{\cal V}}    \def\CW{{\cal W}}    \def\CX{{\cal X}}
\def\CY{{\cal Y}}    \def\CZ{{\cal Z}}
      


\articletitle[]{$2d$ random Dirac fermions: large $N$ approach}


\chaptitlerunninghead{ }

\author{D. Serban}
\affil{Service de Physique Th\'eorique, CE Saclay,
F-91191 Gif-sur-Yvette, France}
\email{serban@spht.saclay.cea.fr}




\begin{abstract}
We study the symmetry classes for the random Dirac fermions in 2
dimensions. We consider $N_f$ species of fermions, coupled
by different types of disorder. We analyse the
renormalisation group flow at the order of one loop. At $N_f$ large,
the disorder distribution flows to an isotropic distribution and
the effective action is a sigma model. 
\end{abstract}



\section{Introduction}

Random Dirac fermions in 2 dimensions appear in
various contexts in condensed matter physics.
They were used to model disorder
in systems with degenerate Fermi surface \cite{fradkin} ,
in the $d$-wave superconductors \cite{ntw,asz} or to investigate the
plateau transition in the
integer quantum Hall effect \cite{lfsg} .
In all these problems, the main question concerns the localization of
quasiparticles, often different from the generic
Anderson localization.

The universal properties of a system where randomness is present
depend largely on the discrete symmetries, like time reversal invariance or
spin rotation symmetry.
This principle governs the Wigner-Dyson classification
of random hermitian  matrices \cite{Dyson}. Considering other discrete
symmetries, like particle-hole or chiral symmetry, led Altland and Zirnbauer
\cite{zirn}
to extend this classification.
However, it was recognized that the physics of 2d Dirac fermions
is richer than expected from the random matrix classification, in the
sense that for the same symmetry class, some systems may be critical and
other not. One reason is that in two dimension,
the sigma models, which are supposed to describe the universal physics,
can support a topological term. In the form of a theta
or WZW term, they can render the model critical
\cite{rgsf,fendley}.
Sometimes, the presence of such a term can be deduced from
general considerations \cite{rgsf,fendley},
in other situations it can be obtained
exactly \cite{ntw,dcft}. A possible strategy, adopted in
\cite{classd,classc}
and applied here to all classes,
is to try to derive the effective sigma model, topological term included, by
chain of controlled approximations. The drawback is that we have to consider
a 
large number of fermion species $N_f$ in order to have a control parameter
for the approximation. This means that the most interesting cases,
where $N_f$ is small, are out of reach of this method.

The symmetry classes for 2d random Dirac hamiltonians, were recently
analysed \cite{denis}. The constraint that the discrete
symmetries should also preserve the form of the
Dirac operator leads to a refinement  of the Altland-Zirnbauer
classification,  in which the classes
$A {\rm  III}$,  $C {\rm  I}$  and $D {\rm  III}$,
appear each in two different forms.
Here, we pursue the characterization of the symmetry classes for
the Dirac Hamiltonians of \cite{denis}.  We compute the
beta function for all the disorder perturbations.
When the number of fermion components $N_f$ is large the flow of
the coupling constants generically closes on a single
coupling constant, a fact  which simplifies considerably
the derivation of the effective action.
This was done for classes $D$ in \cite{classd}
and for classes $C$ and $C {\rm I}$ in \cite{classc}, and we refer to these
papers for the details of the method.
Without surprise, we obtain the sigma models defined
on the manifolds identified by Zirnbauer
\cite{zirn} . The two different versions of classes
$A {\rm  III}$,  $C {\rm  I}$  and $D {\rm  III}$
give rise to sigma models with or without a WZW term.
   
We consider random Dirac Hamiltonians in 2d, in the representation
\begin{equation}
  H = \pmatrix{V+M &-2i\partial_z +A_{ \bar z } \cr
     -2i\partial_{  z }+A_{ \bar z }&V-M\cr}
  \;, \label{hamilt}
\end{equation}
where $\partial_z =(\partial_x-i\partial_y)/2 $,
$\partial_{\bar z }=(\partial_x+i\partial_y)/2 $ and
$V$, $M$, $A_z$ and $A_{ \bar z }$ are random matrix fields
of dimension $N_f$.

The discrete symmetries
preserving the form of the Dirac Hamiltonian   are \cite{denis}
\begin{eqnarray}
 H &=& - P\, H \, P^{-1},\quad
P=\pmatrix{\ga & 0\cr 0& -\ga\cr},\ PP^{\dag}=1,\ P^2=1
\label{Psym}\\
H &=& \ep_c\, C\, H^T \, C^{-1},\quad
C=\pmatrix{0& \sigma \cr  -\ep_c\sigma &0 \cr},\ CC^{\dag}=1,\  C^T=\pm C\;.
\label{Csym}
\end{eqnarray}
The first condition will be called chiral symmetry, while the second one
could be called particle-hole symmetry when $\ep_c=-1$ and time reversal
symmetry (or spin rotation invariance) when $\ep_c=+1$.
The symmetry conditions impose constraints on the matrix
fields $V$, $M$, $A_z$ and $A_{ \bar z }$, which are listed in \cite{denis}.
We consider gaussian
probability distribution for the disorder, compatible with
the symmetry conditions.

The spectral properties of the Hamiltonians \ref{hamilt} can be analysed
using the supersymmetry \cite{efetov} (or replica) method, in which
products of Green functions
are represented by gaussian integral over equal numbers of
fermionic and bosonic Dirac fields. Average over gaussian randomness
introduce effective coupling between the bosonic and fermionic fields,
the coupling constants being the disorder variances.
 
The first step of our analysis is the renormalisation group flow for the
effective coupling constants. The analysis can be
eventually performed at all orders
\cite{leclair} . 
In the section \ref{beta}
we give the one loop beta functions for all the clases and for any $N_f$.
Generically, at large $N_f$ the flow closes
on a single coupling constant, corresponding to equal variances for
different types of disorder.
\footnote{ Exceptions may be the coupling constants
corresponding to some $u(1)$ part of the disorder,
specific to classes $0$, $1$, $2$ and $9_\pm$.
This situation was analysed in the case $1\ (A {\rm III})$ in \cite{asz}
and in \cite{ludwig} in a non-hermitian version of case $0$.
For the moment, we ignore this point and
analyse the case of equal disorder variances.
}
In this case, the effective interaction, quartic in the
the Dirac fields, can be decoupled via a Hubbard-Stratonovich
transformation using a (super)matrix field $Q$ whose dimension does not
depend on $N_f$. The number of components of the Dirac fields appear
as an overall factor and it can be used as control parameter
for a saddle point approximation.

The effective action for the field $Q$, hence
the saddle point equation, has global invariance, described by a group
$G$. The solutions of the saddle point equation break the invariance
to a subgroup $H$, so that the manifold of saddle points
is given by $G/H$.
The low energy behaviour is obtained by taking
slowly varying configurations of the field on $G/H$
(Goldstone modes) and evaluating the action
for these modes. 
The result is a sigma model on some symmetric
(super) space. Given the fact that in two dimensions sigma models
can support a topological term, we may obtain a theta term or a
WZW one. 
In the last section, we give a list of bilinears in Dirac fields
to which couple the field $Q$, the invariance
group $G$ of $Q$, as well as a
diagonal solution of the saddle point equation $q_0$ together with its
stabiliser $H$. Together with the coupling constants,
these are the basic ingredients to characterize the sigma model
for each symmetry class.
 
\section{Disorder perturbations}

We use notations similar to \cite{classd} .
To obtain a compact notation, we introduce supermultiplets
\begin{equation}
\label{multi}
\phi = \pmatrix{\psi_+\cr \psi_-^T\cr  \be_+ \cr  \be_-^T \cr} \;,
\qquad  \phi^{ t} \equiv ( \psi_- ,\; \psi_+^T , \; \beta_-  ,\; - \beta_+^T
)
\end{equation}
and similarly for $\bar \phi$, $\bar \phi^{ t}$, where $\psi_+$ and $\psi_-$
are fermion multiplets with $N_f$ components  while $\beta_+$ and $\beta_-$
are 
bosonic ghostswith the same number of components. The superscript $T$
denotes
transposition. 

In the absence of disorder,
the action for the Dirac fields is written
\begin{equation}
\label{free}
S_{{\rm cft}}=\int \frac{d^2x}{2\pi} (\phi^{ t} \partial_{\bar z}
\phi + \bar \phi^{ t} \partial_{ z}\bar \phi)\;,
\end{equation}
and it corresponds to a conformal field theory with central charge
$c=0$. From the holomorphic (antiholomorphic) fields
$\phi$ $(\bar \phi)$, one can construct the current algebra
$OSp(2N_f|2N_f)_{k=1}$ \cite{dcft} .

The perturbation induced by disorder is
\begin{equation}
S_{{\rm pert}}=g_i \int \frac{d^2x}{2\pi}\CO_i\;,
\end{equation}
where $g_i$ are  the disorder variances associated to different
disorder invariants
and the operators $\CO_i$ are defined as follows
\begin{eqnarray}
\CO_A&=&\sum_{I\in \CA} (\phi^t E^-_I \phi)( \bar \phi^t E^-_I \bar \phi)
\;,
\qquad \CO_a= (\phi^t E^-_0 \phi)( \bar \phi^t E^-_0 \bar \phi) \;,\cr
\CO_V&=&\sum_{I\in \CV} (\bar \phi^t E^-_I \phi)^2\;,
\qquad \CO_v=(\bar \phi^t E^-_0 \phi)^2\;,\cr
\CO_M&=&\sum_{I\in \CM} (\bar \phi^t E^+_I \phi)^2\;,
\qquad \CO_m=(\bar \phi^t E^+_0 \phi)^2\;,
\end{eqnarray}
where 
$$E_I^{\pm}=\pmatrix{e_I & 0\cr 0 & \pm e_I^T}\;,$$
$e_I$ are generators of $gl(N_f)$
normalized by ${\rm Tr} (e_I e_J)=\de_{IJ}$
and $\CA,\ \CV,\ \CM$
are subsets of indices of $sl(N_f)$ which depend on the symmetry
class.
$e_0$ is the identity normalized by ${\rm Tr}\; e_0^2=1$.

\section{The one loop beta functions}
\label{beta}

To study the renormalisation group flow for the coupling constants,
we need the corresponding beta function. They can be derived
from the operator product expansion of the perturbing operators
$\CO_i$ 
\begin{equation} 
{\cal{O}}_i(z,\bar{z}) {\cal{O}}_j(0)
\simeq\frac{1}{z\bar{z}} C^k_{ij} {\cal{O}}_k(0)+ {\rm reg.} \nonumber
\end{equation}
The one loop beta functions are given by \cite{cardy}
\begin{equation} 
\beta_k\equiv l\d_l\, g_k =-\sum_{i,j} C^k_{ij} g_i g_j
\nonumber
\end{equation}
The results for different classes are listed in the
remaining of this section.

{\bf \large{The non-chiral classes}}

{\bf Class $0$ ($A$)}

This class do not posses neither chiral invariance, nor particle-hole/
time reversal symmetry.We set $N_f=N$
and consider only disorder which do not break parity invariance of the
action.
The disorder fields $A$, $V$ and $M$ belong to $gl(N)$; the $sl(N)$
part generates the perturbation operators $\CO_A,\ \CO_V,\ \CO_M$,
while the $gl(1)$ part generates $\CO_a,\ \CO_v,\ \CO_m$.
The corresponding beta
functions are
\begin{eqnarray*}
\be_A &=& -\frac{D}{2} g_A^2- \frac{D}{8}\(g_M^2+g_V^2\)+\frac{B}{4} g_M g_V
+\frac{1}{N} g_m g_V+\frac{1}{N} g_M g_v\;,\cr
\be_a &=& \frac{N y}{4} g_M g_V + \frac{1}{N} g_m g_v\;,\cr
\be_M &=& 
-
x g_M^2-\frac{1}{N}
(g_m+g_v) g_M +\(2x-\frac{D}{2}\)g_A g_M + \frac{B}{2} g_A g_V\cr
&+&\frac{2}{N} g_A g_v
+\frac{2}{N} g_a (g_V+g_M)-x g_M g_V\;,\cr
\be_V &=& x g_V^2+\frac{1}{N} (g_m+g_v) g_V +\(2x-\frac{D}{2}\)g_A g_V
+ \frac{B}{2} g_A g_M\cr
&+&\frac{2}{N} g_A g_m
+\frac{2}{N} g_a (g_M+g_V)+x g_M g_V\;,\cr
\be_m &=& - 
\frac{1}{N} g_m(g_m+g_v)-C g_m (g_M+g_V)+2C g_A g_m \cr
&+&
\frac{2}{N} g_a (g_m+g_v)
+\frac{Ny}{2} g_A g_V
\;,\cr
\be_v &=&  \frac{1}{N} g_v(g_v+g_m)+C g_v ( g_V+g_M) +2C g_A g_v \cr
&+&
\frac{2}{N}g_a (g_v+g_m)
+ \frac{Ny}{2} g_A g_M
\;,
\end{eqnarray*}
where the constants, with group theoretical significance, are
\begin{eqnarray*}
D&=&-2N\;, \quad C=\frac{N^2-1}{N}\;, \quad B=\frac{2(N^2-4)}{N}\;,\cr
x&=&-\frac{1}{N}\;, \qquad y=\frac{4(N^2-1)}{N^2}\;.
\end{eqnarray*}

In the large $N$ limit, the line $ g_A=g_V=
g_M= g_m=-g_v=2g_a= g\sim 1/N $ is
preserved by the RG flow. It is interesting to note that
we cannot obtain an invariant line with all the coupling constants positive.

{\bf Classes $3_-$ and $4_-$ ($D$ and $C$)}

These classes have particle-hole symmetry ($\ep_c=-1$), inforced by
symmetric,
respectively antisymmetric matrix $\sig^T=\pm \sig$. We set $N_f=2N$, but
for class $D$, $N$ can be half integer.
The random fields $A$ and $V$ belong to the algebra $so(2N)$ (case $D$) and
$sp(2N)$ (class $C$). These algebra are spanned by a subset of the
$sl(2N)$ generators $e_I$, $\tau_I$, selected by the condition $\tau_I=-\sig
\tau_I^T\sig^{-1}$.
The field $M$ belongs to the complement of these algebras
with respect to $sl(2N)$, spanned by $T_I$ with $T_I=\sig
T_I^T\sig^{-1}$. Operators  $\CO_a,\ \CO_v$ are not allowed by these
symmetry classes.
The beta functions are
\begin{eqnarray*}
\be_A &=& -\frac{D_{\tau \tau }}{2} g_A^2- \frac{D_{TT}}{8}g_M^2-
\frac{D_{\tau \tau }}{8} g_V^2+\frac{B_{\tau T}}{4} g_M g_V
+\frac{1}{2N} g_m g_V;,\cr
\be_M &=& -x_{TTT} g_M^2-\frac{1}{2N} g_m g_M +\(2x_{\tau T \tau}-
\frac{D_{\tau T}}{2}\)g_A g_M +
\frac{B_{\tau \tau }}{2}
g_A g_V \cr &-& x_{\tau T \tau} g_M g_V\;,\cr
\be_V &=& x_{\tau \tau \tau} g_V^2+\frac{1}{2N} g_m g_V +\(2x_{\tau \tau
\tau}-
\frac{D_{\tau \tau}}{2}\)
g_A g_V + \frac{B_{\tau T}}{2}
g_A g_M\cr &+& \frac{1}{N} g_m g_A
 + 
x_{T \tau T} g_M g_V\;,\cr
\be_m &=&-\frac{1}{2N} g_m^2 +2C g_A g_m +{N y_{\tau \tau}}
g_A g_V-C g_V g_m 
-C' g_m g_M\;,
\end{eqnarray*}
with the group coefficients (again, the upper sign is for class
$D$ and the lower one for class
$C$)~:
\begin{eqnarray*}
C&=&\frac{2N\mp1}{2}\;, \quad
C'=C_{sl(2N)}-C=\frac{(2N\mp1)(N\pm1)}{2N}\;,\cr
D_{\tau \tau}&=&-2(N\mp1)\;, \quad D_{\tau T}=-2N\;, \quad D_{T T}=-
2(N\pm1)\;,\cr
x_{\tau \tau \tau}&=& -x_{\tau T \tau}=\pm\frac{1}{2}\;, \quad
\quad x_{TTT} =-\frac{1\mp N}{2N}\;, \quad
x_{T\tau T} = \frac{1\pm N}{2N}\;,\cr
y_{\tau \tau}&=&\frac{(2N\mp1)}{N}\;, \quad
B_{\tau \tau}=2(N\mp1)\;, \cr
B_{\tau T}&=&2\frac{N^2-1}{N}\;,
\quad B_{T T}=2\frac{(N\pm2)(N\mp1)}{N}\;.
\end{eqnarray*}

At large $N$, 
an invariant line is given by $g_A=g_M=g_V=g_m=g $.

{\bf Classes $3_+$ and $4_+$ ($AII$ and $AI$) }

These classes have time reversal (spin rotation) symmetry,
with $\ep_c=1$ and $\sig^T=\pm \sig$.
They are similar to classes $3_-$ and $4_-$, with the random fields $V$ and
$M$
changing roles: now $M$ belongs to $so(2N)$ and $sp(2N)$ respectively, while
$V$ belongs to the complement with respect to $sl(2N)$.
The beta functions can be obtained from that of the classes  $3_-$ and
$4_-$ by changing $g_M \leftrightarrow -g_V$ and $g_m \leftrightarrow -g_v$.

{\bf \large{The chiral classes}}

$\bullet$ The first type of chiral classes corresponds to $\gamma=1$ in eq.
\ref{Psym}
and it comprises three classes.
Only disorder of the vector potential type is allowed, and these
classes can be 
seen as limits of the previous ones when $g_V,\ g_M,\ g_v,\ g_m\ \to 0$.

{\bf Class $1$ ($AIII$)}

The random gauge potential $A$ belongs to $gl(N)\simeq gl(1)\otimes sl(N)$
($N_f=N$).
\begin{equation}
\be_A = -\frac{D}{2} g_A^2\;, \qquad \be_a=0\;,
\end{equation} 
with $D=-2N$.

{\bf Classes $5$ and $6$ ($DIII$ and $CI$) }

In addition to the chiral symmetry, these classes posses
particle-hole symmetry. We set $N_f=2N$.
The random gauge potential $A$ belongs to $so(2N)$ and $sp(2N)$ respectively
\begin{equation}
\be_A = -\frac{D_{\tau \tau }}{2} g_A^2\;,
\end{equation}
with $D_{\tau \tau }=-2(N\mp 1)$.

$\bullet$ The second type of chiral classes correspond to $\gamma=\sig_3$
and it comprises five classes.
 We set $N_f=2N$, but for classes $8$
and $9_+$ $N$ should be understood as even integer.

{\bf Class $2$ ($AIII$)}

The chiral symmetry imposes the constraint $[A,\gamma]=0$ to the
random gauge potential; this condition selects an algebra
$gl(N)\otimes gl(N)\simeq (gl(1)\otimes sl(N))^{\otimes 2}$ out of $gl(2N)$.
The two $gl(1)$ parts are generated by the identity and $\ga$
and we denote the associated coupling constants by $g_a$ and $g_{a-}$.
We denote by $e_i$ the generators of $sl(N)\otimes sl(N)$.
The two $sl(N)$ components can have different variances, $g_{A1}$ and
$g_{A2}$,
with $g_A^\pm = (g_{A1}\pm g_{A2})/2$.

The fields $V$ and $M$ are constrained by
$\{V,\ga\}=0$, $\{M,\ga\}=0 $, so they belong to
$gl(2N)\setminus (gl(N)\otimes gl(N))$, generated
by $e_\alpha$. Operators $\CO_v$ and $\CO_m$
are not allowed by the chiral classes.

The beta functions are
\begin{eqnarray*}
\be_A^+ &=& -\frac{D_{ii}}{2} (g_A^{+2}+g_A^{-2})
- \frac{D_{\al \al}}{8}\(g_M^2+g_V^2\)
+\frac{B_{\al\al}}{4} g_M g_V\;, \cr
\be_A^-&=&-D_{ii}g_A^+g_A^-\;, \quad
\be_a = \frac{N y_\al}{2} g_M g_V \;,\quad
 \be_{a-}=
- \frac{D_{\al \al}}{8}\(g_M^2+g_V^2\)\;,\cr
\be_M &=& -x_{\al \al \al} g_M(g_M+g_V)-\frac{D_{i\al}}{2}g_A^+ g_M
+ \frac{B_{i\al}}{2} g_A^+ g_V+\frac{1}{N} g_a (g_M+g_V)\;,\cr
\be_V &=& x_{\al \al \al} g_V(g_V+g_M) -\frac{D_{i\al}}{2}g_A^+ g_V
+ \frac{B_{i\al}}{2} g_A^+ g_M+ \frac{1}{N} g_a (g_V+g_M)\;.
\end{eqnarray*}
with the group coefficients
\begin{eqnarray*}
D_{ii}&=&D_{sl(N)}=-2N\;,\qquad D_{\al \al}=
D_{gl(2N)}-D_{gl(N)}=-2N\;, \cr
D_{i \al}&=&D_{gl(2N)}/2+\frac{2}{N}=-2N+\frac{2}{N}\;, \qquad
y_\al=2\;, \cr
x_{\al\al\al}&=&0\;, \qquad B_{i\al}=\frac{2(N^2-1)}{N}\;,
\qquad B_{\al \al}=2N .
\end{eqnarray*}

At large $N$, the line $g_A^-=0$, $g_A^+=g_M=g_V=2g_a=4 g_{a-}$
is left invariant by the RG flow.

{\bf Classes $7$ and $8$ ($DIII$ and $CI$) }

These classes have both chiral symmetry and particle-hole symmetry, with
the matrices inforcing these symmetries commuting.

We split the generators of $gl(2N)$ in four groups $\tau_i,\ \tau_\al,
\ T_i,\ T_\al$, with  $\tau_i,\ T_i$ commuting with $\sig_3$
and $\tau_\al,\ T_\al$ anticommuting with it; also,
$\tau_{i,\al}= -\sig \tau^T_{i,\al} \sig^{-1}$ and
$T_{i,\al}= \sig T_{i,\al}^T \sig^{-1}$.
The random gauge potential $A$ is an element of $so(N)\otimes so(N)$
or $sp(N)\otimes sp(N)$generated by $\tau_i$.
If we choose the representation $\ep_c=-1$,
$V$ belongs to $o(2N)\setminus o(N)\otimes o(N)$
or $sp(2N)\setminus sp(N)\otimes sp(N)$ respectively, generated by
$\tau_\alpha$, while the field $M$ is generated by $T_\al$.

The beta functions are
\begin{eqnarray*}
\be_A^-&=&-D_{\tau\tau}^{ii}g_A^+g_A^-\;, \cr
\be_A^+ &=& -\frac{D_{\tau\tau}^{ii}}{2} (g_A^{+2}+g_A^{-2}) -
\frac{D_{T T}^{\al \al}}{8} g_M^2-\frac{D_{\tau\tau}^{\al \al}}{8}g_V^2+
\frac{B_{\tau T}^{\al \al}}{4} g_M g_V\;,  \cr
\be_M &=&-x_{TTT}^{\al \al\al} g_M^2 +\(2x_{\tau T \tau}^{i\al i}-
\frac{D_{\tau T}^{i\al}}{2}\) g_A^+ g_M+ \frac{B_{\tau\tau}^{i \al}}{2}
g_A^+ g_V -x_{\tau T \tau}^{\al \al \al}g_M g_V\;, \cr
\be_V &=&x_{\tau \tau\tau}^{\al \al\al} g_V^2 +\(2x_{\tau \tau\tau}^{i\al
i}-
\frac{D_{\tau \tau}^{i\al}}{2}\) g_A^+ g_V+ \frac{B_{\tau T}^{i \al}}{2}
g_A^+ g_M +x_{ T \tau T}^{\al \al \al}g_M g_V\;,
\end{eqnarray*}
where the group coefficients are
\begin{eqnarray*}
D^{ii}_{\tau\tau}&=&-N\pm 2\;,
\quad D^{\al \al}_{\tau\tau}= D^{\al \al}_{TT}=-N\;, \cr
D^{i \al}_{\tau\tau}&=& D^{i \al}_{\tau T}=   -N\pm 1\;, \cr
x_{TTT}^{\al \al \al}&=&  x_{\tau \tau\tau}^{\al \al \al} =
 - x_{\tau T \tau}^{\al \al \al}
=-x_{T\tau T}^{\al \al \al}=
\pm\frac{1}{2}
\;,\quad  x_{\tau T \tau}^{i \al i}= x_{\tau \tau \tau}^{i \al i}=0\;, \cr
B^{\al \al}_{\tau T}&=&N \;, \quad  B^{i\al}_{\tau \tau}= B^{i\al}_{\tau T}=
N\mp 1  \;.
\end{eqnarray*}
In the large $N$ limit
the RG flow preserve the line $g_A^-=0$, $g_A^+=g_M=g_V$.

{\bf Classes $9_+$ and $9_-$ ($D \rm{I}$ and $C \rm{II}$) }

These classes have also both particle-hole and chiral symmetry, with
the matrices enforcing these constraints anticommuting, $\{\gamma,
\sig\}=0$. 

If we choose $\ep_c=-1$, the matrix $\sig$ defining the particle-hole
symmetry
is symmetric for class $D \rm{I}$
and antisymmetric  for class $C \rm{II}$, as for the cases $3_-$ and $4_-$
or
$7$ and $8$. The opposite choice exchanges the symmetry of the matrix
$\sigma$
between the two classes.

As in the previous case, we split the $gl(2N)$ generators into four classes.
Since $\{\gamma, \sig\}=0$, $\tau_i$
spans now $gl(N)$, to which the field $A$ belong.
$V$ is generated by $\tau_\al$, which span
$o(2N)\setminus gl(N)$ and $sp(2N)\setminus gl(N)$ respectively.
$M$ is again generated by $T_\al$.
\begin{eqnarray*}
\beta_{a-}&=&- \frac{D_{T T}^{\al \al}}{8} g_M^2
-\frac{D_{\tau\tau}^{\al \al}}{8}g_V^2\;, \quad
\be_A^-=-D_{\tau\tau}^{ii}g_A^+g_A^-\;, \cr
\be_A^+ &=& -\frac{D_{\tau\tau}^{ii}}{2} (g_A^{+2}+g_A^{-2}) -
\frac{D_{T T}^{\al \al}}{8} g_M^2-\frac{D_{\tau\tau}^{\al \al}}{8}g_V^2+
\frac{B_{\tau T}^{\al \al}}{4} g_M g_V\;,  \cr
\be_M &=&-x_{TTT}^{\al \al\al} g_M^2 +\(2x_{\tau T \tau}^{i\al i}-
\frac{D_{\tau T}^{i\al}}{2}\) g_A^+ g_M+ \frac{B_{\tau\tau}^{i \al}}{2}
g_A^+ g_V -x_{\tau T \tau}^{\al \al \al}g_M g_V\;, \cr
\be_V &=&x_{\tau \tau\tau}^{\al \al\al} g_V^2 +\(2x_{\tau \tau\tau}^{i\al
i}-
\frac{D_{\tau \tau}^{i\al}}{2}\) g_A^+ g_V+ \frac{B_{\tau T}^{i \al}}{2}
g_A^+ g_M +x_{ T \tau T}^{\al \al \al}g_M g_V\;,
\end{eqnarray*}
with
\begin{eqnarray*}
D^{ii}_{\tau\tau}&=&-N\;,
\quad D^{\al \al}_{\tau\tau}=-N\pm 2\;, \quad    D^{\al \al}_{TT}=-N\mp2\;,
\cr
D^{i \al}_{\tau\tau}&=&    -\frac{(N\pm 1)(N\mp2)}{N} \;,
 \quad   D^{i \al}_{\tau T}=
-\frac{(N\mp 1)(N\pm2)}{N}\;, \cr
x_{TTT}^{\al \al \al}&=&  x_{\tau \tau\tau}^{\al \al \al} =
  x_{\tau T \tau}^{\al \al \al}
=x_{T\tau T}^{\al \al \al}=0
\;, \quad  x_{\tau T \tau}^{i \al i}=\frac{1}{2N}\mp \frac{1}{2}\;,\cr
\quad
x_{\tau \tau \tau}^{i \al i}&=&\frac{1}{2N}\pm \frac{1}{2}\;, \quad
B^{\al \al}_{\tau T}=N \;, \quad  B^{i\al}_{\tau \tau}=N\mp1\;,
\quad B^{i\al}_{\tau T}=N\pm1
  \;.
\end{eqnarray*}

At large $N$
the RG flow preserves the line $g_A^-=0$ and $ g_A^+=g_M=
g_V=4g_{a-}$.

We note that the equations listed above have interesting properties
under the transformation $g_M \leftrightarrow -g_V$ and
$g_m \leftrightarrow -g_v$.
This operation leave the equations for classes $0$, $2$, $7$ and $8$
invariant and exchanges the ones for classes $3_\pm$, $4_\pm$ and $9_\pm$,
therefore it is equivalent to changing the sign of $\ep_c$, when defined
\footnote{We thank to D. Bernard for pointing out that this can be realized
on the Dirac Hamiltonian via a non-unitary transformation.}.
Moreover, on the line $g_M=-g_V=g$ ($=g_m=-g_v$, when they exist), the flow
for $g$ and the flow for $g_A$ decouple. This was noted
in \cite{ludwig} for class $0$ and was related to
"spin-charge" separation,
for class $C$ in \cite{da} .

\section{Hubbard-Stratonovich transformation and the sigma model}

We consider products of $n$ Green functions, $n_R$ retarded and
$n_A$ advanced, and we need to introduce a copy of the
multiplets \ref{multi} for each Green function.
When the Hamiltonian has chiral
or particle-hole invariance, the spectrum is symmetric
with respect to $E=0$ and we do not need to distinguish
between retarded and advanced sector. In this case, the density of
states can be singular at the point $E=0$.

As already mentioned, when the coupling constant have equal values,
the decoupling of the quartic interaction is particularly simple.
At that particular point, the disorder perturbation can be written
only in terms of singlets of $gl(N_f)$.
The decoupling matrix $Q$ couples to these singlets and it has a size
independent of $N_f$.
Up to an additive constant, the effective action for $Q$ is
    \begin{equation}
        S[Q] = - N_f\left[
        {1 \over \tilde g}\int \frac{d^2x}{2\pi} \,{\rm STr}\,Q^2 -
\, 
        {\bf STr}\ln\pmatrix{ Q &\partial\cr \bar\partial &Q\cr}\;,\right]
        \label{action}
        \end{equation}
where $\tilde g \sim N_f g$.
Part of the invariance of the  free action \ref{free} ,
$OSp(2nN_f|2nN_f)$,
survives in the effective action
\ref{action}. We denote the residual invariance group by $G$.
 
The saddle point equation $\de S[Q]/\de Q=0$ can be solved
making a diagonal ansaz
$Q_0=\mu\; q_0$, where $q_0^2=1$ and $\mu$ has the dimension
of a mass. As there is no scale in the problem, $g_i$ being dimensionless,
$\mu$ is dynamically generated.
The group $G$ acts on $Q$ by conjugation, and from a particular
saddle point solution $q_0$ one generates a whole manifold by
$q_0\ \to \ q=g q_0 g^{-1}$. This manifold is isomorphic to $G/H$, where
$H$ is the stabilizer of  $q_0$.
The last step in deriving the sigma model is to evaluate the action
\ref{action} on slowly varying configurations $q(x)$ on $G/H$.
An elegant way of doing it is to use non-abelian bosonisation
\cite {witten}, 
as in \cite{asz,classd} and to write the free action
as a WZW model . The field $q(x) $ becomes
a mass term for the WZW field, which is  forced to follow it.
At scales larger than $\mu^{-1}$, the effective action
is  
    \begin{equation}
        S[q] = -\frac{1}{16\pi f}
        \int \frac{d^2x}{2\pi} \,{\rm STr}\, \partial_\nu q
        \partial_\nu q \; + S_{\rm top}[q]
        \end{equation}
In the following, we give the main characteristics of the
sigma model manifold for the 13 classes.

{\bf Class 0 ($A$}
\begin{eqnarray*}
&Q &\sim  {\rm Tr_{gl(N)}}\;
 [\Sig_3, \phi \bar \phi^t +\bar\phi  \phi^t]\;,
 \quad q_0 = \Sig_3 \otimes \Lambda\;,\cr
&G&=GL(n|n)\;,\quad
H=GL(n_R|n_R)\otimes GL(n_A|n_A)\;,
\end{eqnarray*}
with $\Lambda={\rm diag}(1_{n_R},-1_{n_A})$.

{\bf Classes $3_-$ and $4_-$ ($D$ and $C$)}
\begin{eqnarray*}
&Q&
\sim  {\rm Tr_{gl(2N)}} \;\Sig
\(\phi \bar \phi^t +\bar\phi  \phi^t\)\Sig^{-1}\;,
\quad q_0 = \Sig_3\;, \cr
&G&=OSp(2n|2n)\;,\quad
H=GL(n|n)\;,
\end{eqnarray*}
with $\Sig \equiv {\rm diag} (1, \sig)$

{\bf Classes $3_+$ and $4_+$ ($A {\rm II}$ and $A {\rm I}$) }
\begin{eqnarray*}
&Q&\sim  {\rm Tr_{gl(2N)}} \;\Sig \(\Sig_3\phi \bar \phi^t +\bar\phi  \phi^t
\Sig_3\)\Sig^{-1}\;,\quad q_0 = \Lambda\;,
\cr
&G&=OSp(2n|2n)\;, \quad
H=OSp(n_R|n_R)\otimes OSp(n_A|n_A)\;.
\end{eqnarray*}

$\bullet$ The effective action for the classes $1$, $5$ and $6$ is invariant
under multiplication at left/right by holomorphic/antiholomorphic
group elements. This is a sign that the theory is conformally
invariant and indeed, when deriving the sigma model we obtain
a WZW term. 
 
{\bf Class 1 ($A {\rm III}$)}
\begin{eqnarray*}
&Q&\sim {\rm Tr_{gl(N)}}\; [\Sig_3,\phi \bar \phi^t ] \;,
\quad \bar Q\sim  1/2\;{\rm Tr_{gl(N)}}\; [\Sig_3, \bar \phi \phi^t ]\;,
\cr
&q_0 &= \Sig_3 \otimes \Lambda\;,
\quad G=GL(n|n)_L \otimes GL(n|n)_R \;,\quad
H= GL(n|n)\;.
\end{eqnarray*}
The level of the WZW action is $k=N_f$.
 
{\bf Classes $5$ and $6$ ($D {\rm III}$ and $C {\rm I}$)}
\begin{eqnarray*}
&Q&\sim \; {\rm Tr_{gl(2N)}} \;\(\Sig \phi \bar \phi^t \Sig^{-1}\)
\;,
\quad \bar Q\sim \; {\rm Tr_{gl(2N)}} \;
\(\Sig \bar \phi \phi^t \Sig^{-1}\)\;,
\cr
&q_0& = \Sig_3 \;, 
\quad G=OSp(n|n)_L \otimes OSp(n|n)_R \;,\quad
H= OSp(n|n)\;.
\end{eqnarray*}
The level of the WZW action is $k=\pm 2N$, the change of sign
indicating that the compact and non-compact sector switch
from the orthogonal to the symplectic sector from one class to another.

$\bullet$ To take into account the chiral symmetry for the last five
classes,
we double the supermultiplets by
\begin{eqnarray*} 
\phi\ &\to& \ \Phi =\frac{1}{\sqrt{2}}
\pmatrix{\phi \cr \ga \phi}\;, \quad
\phi^t\ \to\ \Phi^t=\frac{1}{\sqrt{2}}
\pmatrix{\phi^t & \phi^t \ga }\;, \cr
\bar \phi\ &\to& \ \bar \Phi =\frac{1}{\sqrt{2}}
\pmatrix{\bar \phi \cr -\ga \bar \phi}\;,
\quad \bar \phi^t\ \to\ \bar  \Phi^t=\frac{1}{\sqrt{2}}
\pmatrix{\bar \phi^t & -\bar \phi^t \ga }
\end{eqnarray*}

{\bf Class $2$ ($A {\rm III}$)}
\begin{eqnarray*}
&Q & \sim {\rm Tr_{gl(2N)}}\;
 [\Sig_3, \Phi \bar \Phi^t +\bar\Phi  \Phi^t]\;,
 \quad q_0 = \Sig_3 \otimes \sigma_3\;,\cr
&G&=GL(n|n)\otimes GL(n|n) \;,\quad
H=GL(n|n)\;,
\end{eqnarray*}
where $\sigma_3$ is the Pauli matrix acting in the extra space
introduced to take into account the chiral symmetry.
$G$ is selected from $GL(2n|2n)$ by the condition $[g,\sig_1]=0$.

{\bf Classes $7$ and $8$ ($D {\rm III}$ and $C {\rm I}$)}
\begin{eqnarray*}
&Q&
\sim  {\rm Tr_{gl(2N)}} \;\Sig
\(\Phi \bar \Phi^t +\bar\Phi  \Phi^t\)\Sig^{-1}\;,\quad
q_0 = \Sig_3 \otimes \sigma_3 \;, \cr
&G&=OSp(2n|2n)\otimes OSp(2n|2n) \;,\quad
H=OSp(2n|2n)\;.
\end{eqnarray*}$G$ is selected from $OSp(4n|4n)$ by the condition
$[g,\sig_1]=0$

{\bf Classes $9_-$ and $9_+$ ($D {\rm I}$ and $C {\rm II}$)}
\begin{eqnarray*}
&Q&
\sim  {\rm Tr_{gl(2N)}} \;\Sig
\(\Phi \bar \Phi^t +\bar\Phi  \Phi^t\)\Sig^{-1}\;,\quad
q_0 = \Sig_3 \otimes \sigma_3 \;, \cr
&G&=GL(2n|2n) \;,\quad
H=OSp(2n|2n)\;.
\end{eqnarray*}
The group $G$ is selected by from $OSp(4n|4n)$ by
$[g, \Sig_3 \otimes \sig_1]$.

The classes $2$, $7$, $8$ and $9_\pm$ do not allow a theta term; also, there
is 
no WZW term. Note that the sigma model is defined on the same manifold
for the pair of classes $1$ and $2$ (class $A {\rm III}$), $5$ and $7$
(class $D {\rm III}$) and $6$ and $8$ (class $C {\rm I}$); the difference
between the 
two members of a pair consists only on the presence or absence of the WZW
term.

The coupling constant for the sigma model is given by the inverse of the
number of the Dirac fermions, $f=1/N_f$.





%


\end{document}